\documentclass[comunication,amsfonts,amssymb,twocolumn,showpacs]{revtex4}
\usepackage{mathrsfs}
\usepackage{graphicx,hyperref}
\usepackage{bm}

\usepackage{CJK}
\begin{document}

\title{Analytical results for a parity-time symmetric two-level system under synchronous combined modulations}
\author{Xiaobing Luo $^{1,2}$}
\altaffiliation{Author to whom any correspondence should be addressed: xiaobingluo2013@aliyun.com}
\author{Baiyuan Yang $^{1,2}$}
\author{Xiaofei Zhang $^{3}$}
\author{Lei Li $^{1,2}$}
\author{Xiaoguang Yu $^{1,2}$}
\affiliation{$^{1}$ Department of Physics, Jinggangshan University,
Ji'an 343009, China}
\affiliation{$^{2}$ Institute of Atomic, Molecular Physics $\&$ Functional Materials, Jinggangshan University, Ji'an 343009, China}
\affiliation{$^{3}$ Key Laboratory of Time and Frequency Primary Standards, National Time Service Center, Chinese Academy of Sciences, Xi'an 710600, People's Republic of China}
\date{\today}
\begin{abstract}
We propose a simple method of combined synchronous modulations to generate the analytically exact solutions for a parity-time symmetric two-level system. Such exact solutions are expressible
in terms of simple elementary functions and helpful for illuminating some generalizations of appealing concepts originating in the Hermitian system. Some intriguing physical phenomena, such as stabilization of a non-Hermitian system by periodic driving, non-Hermitian analogs of coherent destruction of tunneling (CDT)
and complete
population inversion (CPI), are demonstrated analytically and confirmed numerically. In addition, by using these exact solutions we derive a pulse area theorem for such non-Hermitian CPI in the parity-time symmetric two-level system. Our results may provide an additional possibility for pulse manipulation and coherent
control of the parity-time symmetric two-level system.

\end{abstract}

\maketitle
\section{Introduction}
Two-level system (TLS) has been an important paradigmatic model in many branches of contemporary physics ranging from radiation-matter interactions to  collision physics\cite{Scully,Allen,Shore}. It lies at the heart of modern applications such as quantum control and quantum information processing as well as constitutes the foundation for fundamental studies of quantum mechanics. Coherent manipulation of two-level systems can be achieved and optimized by introducing external fields, e.g., the most commonly used types of periodic\cite{Rabi} and pulse-shaped\cite{Rosen} driving fields. In the past few decades, the two-level quantum system (or a qubit) involving single and/or combined modulations  has proven a
fertile ground for a variety of intriguing phenomena, including self-induced transparency\cite{McCall}, complete
population inversion (CPI)\cite{Rosen}, dynamical stabilization\cite{Bezvershenko}, and coherent destruction of tunneling (CDT)\cite{Grossmann}, to name only a few.

In certain situations, analytically exact solutions of driven two-level problems exist, offering many advantages in developing analytical approaches to the design of qubit control operations. Unfortunately, it is usually extremely difficult
to acquire exactly soluble two-level evolutions so far. Started with the famous examples like the Landau-Zener model\cite{Landau}, Rabi problem\cite{Rabi},
Rose-Zener model\cite{Rosen}, the problems of finding analytic solutions of driven TLSs have been underway for a long time\cite{Bambini}-\cite{Hai}. However, most of the known exact solutions of TLSs are expressed in terms of complicated special functions, such as the Gauss hypergeometric function\cite{Bambini,Hioe,Robinson,Hioe2,Ishkhanyan,Vitanov}, Weber function\cite{Landau,Zener}, associated Legendre function\cite{Simeonov}, Jacobi elliptic function\cite{Gangopadhyay}, and Heun function\cite{Jha,Jha2,Xie,Zhang}. Up to now, only a very few examples of analytically solvable two-level evolutions described only by several of the simplest solutions\cite{Rabi,Hai}, which
renders the control strategies more transparent, have
been reported.

In recent years, parity-time ($\mathcal{PT}$)-symmetric  two-level system have attracted much attention from both theoretical and experimental
studies\cite{Bender,Guo,Ruter}. A surprising finding of such $\mathcal{PT}$-symmetric non-Hermitian Hamiltonian systems is phase transition, where the spectrum
changes from all real to complex when the gain-loss
parameter exceeds
a certain threshold. It has been widely demonstrated that external driving fields can be effectively used for implementing coherent controls of $\mathcal{PT}$symmetry, and extension
of driven two-level systems to non-Hermitian case is currently of broad interest\cite{Moiseyev}-\cite{Luo2}. For the proper control of the $\mathcal{PT}$-symmetric two-level system,
different analytic methods are required.
 For instance, approximate
solutions to the $\mathcal{PT}$-symmetric Rabi model have recently been proposed based on perturbation theory\cite{Lee}. Given the vast literatures about analytical results for Hermitian two-level systems, it would be possible to find a few simple analytical exact solutions to the driven $\mathcal{PT}$-symmetric two-level problems, which will create the possibility of convenient controls of the non-Hermitian systems.

In this paper, we have constructed a set of simple exact analytical solutions for the $\mathcal{PT}$-symmetric two-level systems under synchronous combined modulations, which helps illuminate some generalizations of the appealing concepts which originate in their Hermitian counterparts. In the case of monochromatically synchronous combined modulations, as it is shown, for the nonzero constant modulation (a periodic oscillating driving with a nonzero static component), the driven $\mathcal{PT}$-symmetric two-level system behaves like the non-driven one with a sudden $\mathcal{PT}$ symmetry breaking, whereas for the zero constant modulation (a purely oscillating driving), the driven $\mathcal{PT}$-symmetric two-level system features all-real spectra and a type of ``generalized Rabi oscillations''. Besides, a non-Hermitian analog of coherent destruction of tunneling (CDT), in which the tunneling can be brought to a standstill, has been observed.
For a square sech-shaped synchronous modulation, the exact analytical solutions of simple forms enables us to precisely produce complete
population inversion (CPI) between two states in $\mathcal{PT}$-symmetric systems. The pulse area theorem for such CPI has also been derived analytically and confirmed numerically.  The central purpose of this article is to develop analytical approaches to the design of coherent control of $\mathcal{PT}$-symmetric two-level systems by constructing exact solutions of simple forms through synchronous combined modulations.

\section{Exact solutions and their applications}
\subsection{Model system}
We consider a non-Hermitian version of driven two-level system, in which the equation
of motion for the probability amplitudes $C_1$ and $C_2$ for the two states $|1\rangle$ and $|2\rangle$ is given as
\begin{eqnarray}\label{eq1}
i\frac{d}{dt}\left(
               \begin{array}{c}
                 C_1 \\
                 C_2\\
               \end{array}
             \right)=H(t)\left(
               \begin{array}{c}
                 C_1 \\
                 C_2\\
               \end{array}
             \right)
\end{eqnarray}
with the following Hamiltonian\cite{Joglekar1,Gong,Lee}
\begin{eqnarray}\label{H1}
H(t)=i\gamma(t)\sigma_z-\nu(t)\sigma_x.
\end{eqnarray}
Here $\sigma_x$ and $\sigma_z$ are the usual Pauli matrices, the real Rabi
frequency $\nu(t)$ describes the coupling between the two levels, and $\gamma(t)$
represents a pair of real gain-loss coefficients. The system (\ref{eq1})
can be realized in many physical contexts, such as coupled waveguides with a complex
refractive index\cite{Guo,Ruter}, dissipative
two-state system of cold atom\cite{Li}, and laser-driven atomic or molecular
transitions with decay\cite{Scully,Allen,Shore}. In this work, we focus on how to design simple synchronous combined modulations to give analytic demonstrations of
some intriguing and unexpected physics of non-Hermitian two-level systems. Here synchronous modulations means that the driving terms  $\gamma(t)$ and $v(t)$ have the same time dependence, in other words, the modulation functions $\gamma(t)$ and $\nu(t)$ satisfy the relation $\gamma(t)=R\nu(t)$ with $R$ being a proportional constant. As we
can easily demonstrate, a $\mathcal{PT}$-symmetric two-state problem is obtained if $\gamma(t)$ and $\nu(t)$ (e.g., the cosine and square sech forms) is temporally
symmetric with respect to a certain point of time.

To obtain the exact solutions of model (\ref{eq1}), we eliminate $C_{2}(t)$  and arrive at a second-order linear differential equation for $C_{1}(t)$ from Eq.~(\ref{eq1}),
\begin{equation}\label{SDE}
    \ddot{C_{1}}(t)-\frac{\dot{\nu}(t)}{\nu(t)}\dot{C_{1}}(t)+[\nu(t)^{2}-\gamma(t)^{2}]C_{1}(t)=0.
\end{equation}
In the case of synchronous modulations, $\gamma(t)=R\nu(t)$,
by introducing a new time scale $\tau=\tau(t)=\int\nu(t)dt$, we can express the second-order differential equation of $C_{1}(t)$ in a simple form,
\begin{equation}\label{NSDE}
\frac{d^{2}{C_{1}}}{d\tau^{2}}+(1-R^{2})C_{1}=0.
\end{equation}

The general solution of Eq.~(\ref{NSDE}), mathematically well
known, falls into three cases
as follows:

\emph{Case 1}: when $R<1$, letting $R=\sin\theta,\sqrt{1-R^{2}}=\cos\theta$, the solution of Eq.~(\ref{NSDE}) is given by
\begin{equation}\label{S11}
  C_{1}(t)= D_{1} e^{-i\tau \cos\theta}+D_{2} e^{i\tau \cos\theta}
\end{equation}
with $D_{1}$ and $D_{2}$ being the undetermined constants determined by
the initial conditions.
According to Eq.(1), $C_{2}(t)$ is easily obtained as
\begin{equation}\label{S12}
  C_{2}(t)=-D_{1} e^{-i\theta-i\tau \cos\theta}+D_{2} e^{i\theta+i\tau \cos\theta}.
\end{equation}
\emph{Case 2}: when $R=1$, Eq.~(\ref{NSDE}) gives the analytical solution of Eq.~(\ref{eq1})
as
\begin{eqnarray}\label{S2}
  C_{1}(t)&=& D_{1}\tau+D_{2} \\
  C_{2}(t)&=&i D_{1}\tau-i D_{1}+i D_{2}.\label{S22}
\end{eqnarray}
\emph{Case 3}: when $R>1$, letting $R=\cosh\phi,\sqrt{R^{2}-1}=\sinh\phi$, and combining Eq.~(\ref{NSDE}) with Eq.~(\ref{eq1}), the exact solutions of $C_{1}$ and $C_{2}$ are obtained as
\begin{eqnarray}\label{S3}
 C_{1}(t)&=& D_{1} e^{\tau \sinh\phi}+D_{2} e^{-\tau \sinh\phi}\\
C_{2}(t)&=&i D_{1} e^{-\phi+\tau \sinh\phi}+i D_{2} e^{\phi-\tau \sinh\phi}.\label{S33}
\end{eqnarray}
Any of the synchronous modulation functions $\nu(t)$ and $\gamma(t)$ will produce an analytical solution to the Schr\"{o}dinger
equation (\ref{eq1}), which thus
 enables us to generate an infinite variety of
analytically solvable $\mathcal{PT}$-symmetric two-state problems along with their explicit solutions (\ref{S11})-(\ref{S33}). Here we will display some interesting physics
by taking several specific cases as examples.

\subsection{synchronous modulation of a monochromatically periodic form}
 We present firstly the case of usual periodic modulation,
 \begin{eqnarray}\label{PM1}
 \nu(t)=\nu_{0}+\nu_{1}\cos\omega t,
 \end{eqnarray}
 where the new time scale is given by $\tau(t)=\nu_{0}t+\frac{\nu_{1}}{\omega}\sin\omega t$. If the particle is initially populated at level $|1\rangle$, we have the initial condition
$C_{1}(0)=1, C_{2}(0)=0$.  Applying them to Eqs.~(\ref{S11})-(\ref{S33})
yields the undetermined constants $D_j (j = 1,2)$ and the exact population dynamics as follows

(i) when $R<1$,
\begin{eqnarray}\label{C1R1I1}
    |C_{1}(t)|^{2}
                  &=&\frac{1}{2\cos^{2}\theta}+\frac{\cos(2\tau \cos\theta-2\theta)}{2\cos^{2}\theta},\\
     |C_{2}(t)|^{2}
                  &=&\frac{1}{2\cos^{2}\theta}-\frac{\cos(2\tau \cos\theta)}{2\cos^{2}\theta},
\end{eqnarray}

(ii) when $R=1$,
\begin{eqnarray}\label{C1R2I1}
    |C_{1}(t)|^{2}&=&\tau^{2}+2\tau+1, \\
     |C_{2}(t)|^{2}&=&\tau^{2},
\end{eqnarray}

and (iii) when $R>1$,
\begin{eqnarray}\label{C1R3I1}
    |C_{1}(t)|^{2}
                  &=&\frac{\sinh^{2}(\tau \sinh\phi+\phi)}{\sinh^{2}\phi},\\
     |C_{2}(t)|^{2}
                 &=&\frac{\sinh^{2}(\tau \sinh\phi)}{\sinh^{2}\phi}.\label{C1R3I11}
\end{eqnarray}

\begin{figure}[htbp]
\centering
\includegraphics[width=8cm]{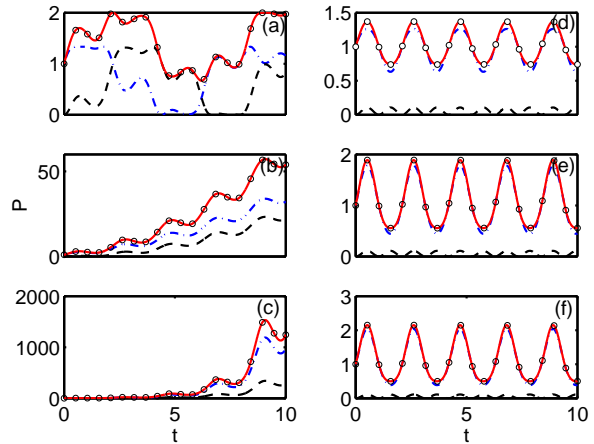}
\caption{(color online) Time evolution of population distributions versus the dimensionless time with $C_{1}(0)=1, C_{2}(0)=0$ for the system (\ref{eq1}) subject to a periodic modulation (\ref{PM1}) with $\nu_0=0.5$ (left column) and  $\nu_0=0$ (right column). From top to bottom: (a) and (d) $R=0.5$; (b) and (e) $R=1$; (c) and (f) $R=1.2$. Shown here are
the numerical results of population $P_1 = |C_1|^2$ in level $|1\rangle$
(blue dashed-dotted line), $P_2 =|C_2|^2$ in level $|2\rangle$ (black dashed
line), and the total population $P = P_1 + P_2 $ (red solid line). The
open circles denote the exact analytical results of the total population.
The other parameters are $\nu_1=1, \omega=3$.} \label{fig1}
\end{figure}

From Eqs.~(\ref{C1R1I1})-(\ref{C1R3I11}), we immediately have the following observations: (1) for zero constant modulation ($\nu_0=0$), the solution should be periodic irrespective of the value of proportional constant $R$. (2) for the nonzero constant modulation ($\nu_0\neq 0$), the solution does
not have to be bounded. Indeed, as long as the proportional constant $R$ is smaller
than the critical value, $R=1$, the system will exhibit oscillations with bound, while as $R$ increases above the critical value, it will exhibit exponential growth. At the critical value, $R=1$, the
time-averaged population shows a parabolic growth. A naive intuition might suggest that apart from a distinct time scale, periodic synchronous driving will bring no new physics to the non-Hermitian system. Indeed, for the nonzero constant modulation ($\nu_0\neq 0$), the driven $\mathcal{PT}$-symmetric two-level system acts like the undriven one, with a phase transition from full to broken $\mathcal{PT}$ symmetry. However, for the zero constant modulation ($\nu_0=0$), stabilization of the non-Hermitian system can be achieved by the periodic synchronous driving in the full range of system parameters, which is contrary to our naive intuition. This is confirmed
by numerical simulations in Fig.~\ref{fig1}.

In Fig.~\ref{fig1}, we show the time evolutions of the populations for both $\nu_0\neq 0$ and  $\nu_0=0$ cases with the initial state $C_{1}(0)=1, C_{2}(0)=0$.
In our simulation, the
numerical solutions of the equation (\ref{eq1}) are
obtained through the use of the Runge-Kutta method.
As expected, for $\nu_0\neq 0$, like the non-driving situation, the system experiences a transition from a state with quasi-periodic oscillation ($\mathcal{PT}$-symmetric state, $R<1$, see Fig.~\ref{fig1}(a)) to a state with unbounded growth (broken $\mathcal{PT}$-symmetric state, $R>1$, see Fig.~\ref{fig1}(c)), whereas for $\nu_0=0$, stable and
coherent but non-norm-preserving population oscillations, referred to as generalized Rabi oscillation in Ref.~\cite{Gong},
are observed in  Fig.~\ref{fig1}(d)-(f) for all values of the proportional constant $R$. We have compared the analytical (circles) with the numerical (red solid lines) results in Fig.~\ref{fig1}. It is clearly seen that they agree well.

As is well known, Floquet theory offers a general framework for treating periodically driven systems. The quasienergies and quasienergy states (the so-called Floquet states), as predicted by Floquet theory, are two basic concepts
for analyzing the periodically
driven system. For the periodic synchronous driving,  we can write the Floquet solution to the time-dependent
Schr\"{o}dinger equation (\ref{eq1}) as $(C_1,C_2)^T=u(t)\exp(-i\varepsilon
t)$, where the Floquet state $u(t)=(\tilde{c}_1,\tilde{c}_2)^T$ is periodic with the modulation
period $\Lambda=2\pi/\omega$, and $\varepsilon$ is the corresponding  quasienergy.

In fact, when one of $D_1$ and $D_2$ equals zero, Eqs.~(\ref{S11})-(\ref{S33}) give the two linearly independent Floquet states (eigenstates) and the corresponding quasienergies as

(i) when $R<1$,
\begin{eqnarray}\label{F1}
   u_{1,2}(t)&=&\left(
            \begin{array}{c}
              \exp[\mp i\nu_{1}\cos\theta\sin\omega t/\omega] \\
              \mp \exp[\mp i(\theta+\nu_{1}\cos\theta\sin\omega t/\omega)]\\
            \end{array}
          \right),\\\varepsilon_{1,2}&=&\pm\nu_{0}\cos\theta, \label{E1}
\end{eqnarray}
and (ii) when $R>1$,
\begin{eqnarray}\label{F2}
   u_{1,2}(t)&=&\left(
            \begin{array}{c}
              \exp[\pm \nu_{1}\sinh\phi\sin\omega t/\omega] \\
               i\exp[\pm (-\phi+\nu_{1}\sinh\phi\sin\omega t/\omega)]\\
            \end{array}
          \right),\\\varepsilon_{1,2}&=&\pm i\nu_{0}\sinh\phi.  \label{E2}
\end{eqnarray}
 According to the above analysis, for the nonzero constant modulation ($\nu_0\neq 0$), there exists a transition between unbroken $\mathcal{PT}$ symmetry (the spectrum is all real, $\varepsilon_{1,2}=\pm\nu_{0}\cos\theta$ if
 $R<1$ ) and broken $\mathcal{PT}$ symmetry (both eigenvalues are imaginary, $\varepsilon_{1,2}=\pm i\nu_{0}\sinh\phi$ if $R>1$), however, for the zero constant modulation ($\nu_0=0$), the two eigenvalues (quasienergies) are degenerate and always equal to zero. Now we take
a closer look at what happens at the transition point, $R=1$.  At this point, for $\nu_0\neq 0$, the two Floquet modes coalesce to $u_{1,2}=(1,i)^T$ and the system has only one linearly independent
eigenstate, leading to an exceptional point. Nevertheless, for $\nu_0=0$, at the point $R=1$, the two linearly independent Floquet modes are given by $u_{1}=(1,i)^T$ and $u_{2}=(\nu_{1}\sin\omega t/\omega, i\nu_{1}\sin\omega t/\omega-i)^T$ respectively, with corresponding degenerate eigenvalues equal to zero.
It should be emphasised that for $\nu_0=0$, the general solution (\ref{S11})-(\ref{S33}) itself is also a kind of Floquet state with a zero quasienergy, due to the fact that any linear combination of two degenerate Floquet eigenstates (the coefficients of linear combination $D_j(j=1,2)$ are adjusted by the initial condition)
is also a Floquet eigenstate of the Hamiltonian operator
corresponding to the same eigenvalue (quasienergy).
\begin{figure}[htbp]
\centering
\includegraphics[width=8cm]{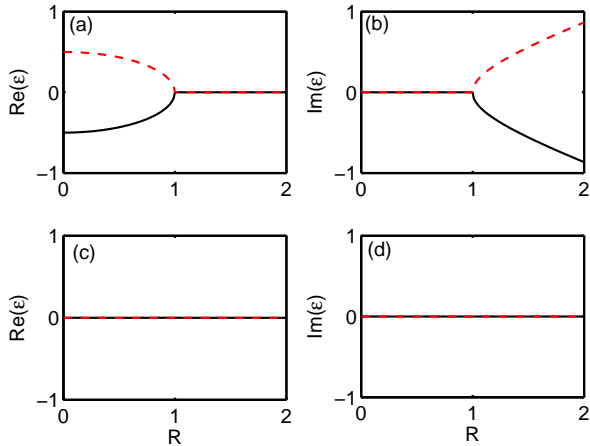}
\caption{(color online)  Real (left) and imaginary (right) parts of the quasienergies
as a function of the proportional constant $R$ for the system (\ref{eq1}) subject to a periodic modulation (\ref{PM1}) with (a)-(b) $\nu_0=0.5$ (upper row) and (c)-(d) $\nu_0=0$ (lower row). The other parameters are $\nu_1=1, \omega=3$.} \label{fig2}
\end{figure}

In Fig.~\ref{fig2}, we exhibit the numerically obtained
quasienergies $\varepsilon$ of the system (\ref{eq1}) for both $\nu_0\neq 0$ and $\nu_0=0$ cases. The quasienergies are numerically computed through direct diagonalization
of the time
evolution operator over one period of the driving, $U(\Lambda, 0) =\mathcal{T} \exp[-i\int_0^\Lambda H(t)dt]$, where $\mathcal{T}$ is the time-ordering operator. As we can see in Figs.~\ref{fig2} (a) and (b), for $\nu_0\neq 0$, the numerical eigenvalue (quasienergy) spectrum indeed changes from being real to complex, indicating a spontaneous $\mathcal{PT}$-symmetry-breaking transition. However, for $\nu_0=0$, the resulting periodically driven two-level system, although being non-Hermitian, processes a pair of degenerate all-real quasienergies ($\varepsilon_1=\varepsilon _2=0$ ) for the whole range of tuning parameters, as illustrated in Figs.~\ref{fig2} (c) and (d). These numerical results are completely consistent with our theoretical analysis.

Coherent destruction of tunneling (CDT) is one of the notable effects of the periodically driven system, a quantum interference effect discovered originally in the periodically driven Hermitian two-level system, upon the occurrence of which the tunneling dynamics can be brought
to a complete standstill. The CDT is connected to the degeneracy
of the quasienergies. In fact, CDT requires in addition that the degenerate
Floquet states do not show
appreciable amplitude oscillations within one
oscillation cycle, which can be
realized in the high frequency limit ($\omega\gg \nu$).
Interestingly, in the non-Hermitian case, CDT still exists if the quasienergies of the underlying time-periodic system are degenerate.

\begin{figure}[htbp]
\centering
\includegraphics[width=8cm]{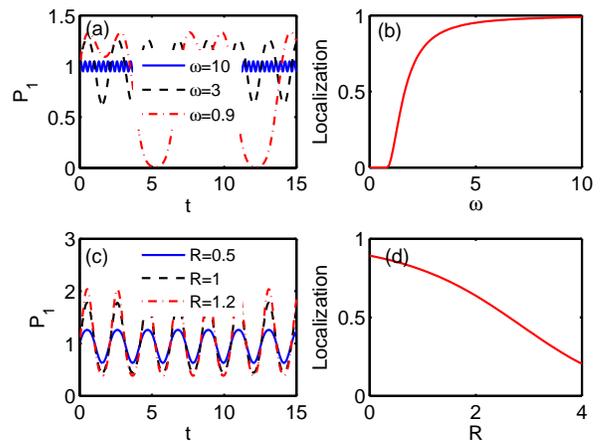}
\caption{(color online) Dynamics of the system (\ref{eq1}) subject to a periodic modulation (\ref{PM1}) with a zero static component $\nu_0=0$, starting from the initial state $C_{1}(0)=1, C_{2}(0)=0$.
(a) Time evolution
of the probability $P_1 = |C_1|^2$, with $\nu_1=1, R=0.5$ and different values of the driving frequency; (b) Localization, defined as the minimum value of $P_1/P$, as a function of $\omega$, with the
same parameters as in (a); (c) Time evolution
of the probability $P_1 = |C_1|^2$, with $\nu_1=1,\omega=3 $ and different values of the proportional constant; (d) Localization as a function of $R$, with the
same parameters as in (c).} \label{fig3}
\end{figure}

Next, we study further the dynamics of the system (\ref{eq1}) for the case of $\nu_0=0$, where the two quasienergies are degenerate and equal to zero.
The results of the numerical simulation
of the probability
distribution $P_1 = |C_1|^2$ are presented in Fig.~\ref{fig3}(a) for different values of driving frequencies. For the low-frequency driving, 
the occupation
of level $|1\rangle$ oscillates and at certain times drops to zero, indicating
no suppression of tunneling. As the driving frequency $\omega$ is increased, the
degree of localization is enhanced and the amplitude of oscillations
of $P_1$ is smaller. If $\omega$ is increased further,
$P_1$ remains near unity and the tunneling dynamics is frozen, which is a direct signature of CDT.
To quantify the stable but non-unitary dynamics more precisely, we have used localization, which is
defined as the minimum value of $P_1/P$, to measure the suppression
of tunneling.  When CDT
occurs, the localization is close to 1; when there is no
suppression, localization is zero.
In  Fig.~\ref{fig3}(b) we plot the minimum value of $P_1/P$, or the localization, as a function of the driving frequency. For small values
of $\omega$, the minimum value of $P_1/P$ equals zero, and thus no CDT occurs. As $\omega$ is
increased, however, the localization is close to one, and the oscillations in $P_1$ are quenched, producing CDT.
We have also examined the effects of the proportional constant $R$ on the localization as shown in Figs.~\ref{fig3}(c) and (d), where with the increase of $R$ the amplitude of oscillations
of $P_1$ is larger and the localization is decreased accordingly.

Since the time scale $\tau$ of the driven system is inversely proportional to the driving frequency, it follows from Eqs.~(\ref{F1})-(\ref{E2}) that the amplitude oscillations of the degenerate
Floquet doublet would be vanishingly small in the high frequency limit. So far, we have demonstrated the existence of CDT in the non-Hermitian system, for which the basic two conditions, degeneracy of quasienergies and absence of appreciable amplitude oscillations of the two degenerate
Floquet states, must be simultaneously
satisfied. This is in full analogy to the Hermitian case.

\subsection{synchronous modulation of a square sech-shaped form}
We now consider the synchronous modulation of a square sech-shaped (a pulse-shaped) form. Let
\begin{eqnarray}
\nu (t)=A \textrm{sech}^{2} t,\label{M2}
\end{eqnarray}
such that the new time variable becomes
\begin{eqnarray}\label{NT2}
\tau(t)=A \tanh t.
\end{eqnarray}
When $R=0$ (hence $\gamma (t)=R\nu (t)=0$), the system (\ref{eq1})
reduces to the famous Rose-Zener model without detuning\cite{Rosen}. We first briefly overview the main results of the Rose-Zener model on resonance.
Therein the transition probability is given as
\begin{eqnarray}\label{TP1}
p=\sin^2 S, ~~S=\int_{-\infty}^{\infty}\nu (t)dt=2A.
\end{eqnarray}
Here $S$ is the area of pulse. Obviously, if the pulse area satisfies  $S=2A=n\pi,n=1,2,3,...$, this transition probability vanishes and the system completely returns its original state, while if $S=2A=(n+\frac{1}{2})\pi,n=0,1,2,3,...$, this pulse produces complete
population inversion (CPI) between two states. Note that the definition
of pulse area used here  differs by a factor 2 from the common usage in the literatures\cite{Rosen,Hioe,Vitanov}.
Surprisingly, we find that there still exist CPI in the driven non-Hermitian ($\mathcal{PT}$-symmetric) two-level system. Hence, below we shall discuss
non-Hermitian generalization of CPI and formulate the
pulse area theorem for such CPI based on the exact solutions (\ref{S11})-(\ref{S33}).

Assuming the initial condition
\begin{eqnarray}\label{IS1}
C_{1}(-\infty)=0,~~C_{2}(-\infty)=1,
\end{eqnarray}
i.e., the system is in the level $|2\rangle$ at $t=-\infty$, we look at the long-time properties of the solution and the final level population at $t=+\infty$. Three distinct cases for this problem are listed as follows.

\emph{The case $R<1$.} Inserting the initial condition (\ref{IS1}) into Eqs.~(\ref{S11})-(\ref{S12}) and (\ref{NT2}) gives
\begin{eqnarray}\label{UC1}
D_{1}=-\frac{e^{-i A \cos\theta}}{2\cos\theta},~~D_{2}=\frac{e^{i A \cos\theta}}{2\cos\theta}.
\end{eqnarray}
The final level populations at $t=+\infty$, according to the exact solutions (\ref{S11})-(\ref{S12}), are thus given by
\begin{eqnarray}\label{C2S1NI2}
     |C_{1}(+\infty)|^{2}&=\frac{\sin^{2}(2A\cos\theta)}{\cos^{2}\theta},\\
    |C_{2}(+\infty)|^{2}&=\frac{\cos^{2}(2A\cos\theta+\theta)}{\cos^{2}\theta}.\label{C2S1NI22}
\end{eqnarray}
From Eqs.~(\ref{C2S1NI2}) and (\ref{C2S1NI22}), we readily observe the following circumstances:

(1) If
\begin{eqnarray}\label{APT1}
 2A\cos\theta=n\pi, n=1,2,3,...,
\end{eqnarray}
then the population restores to its initial state at $t=+\infty$;

(2) If
\begin{eqnarray}\label{APT11}
 2A\cos\theta+\theta=(n+\frac{1}{2})\pi, n=0,1,2,3,...,
\end{eqnarray}
then the population is completely inverted to the level $|1\rangle$ at $t=+\infty$.

It is clearly seen that when $R=0$ (hence $\cos\theta$=1), Eqs.~(\ref{APT1}) and (\ref{APT11}) recover the characteristic of pulse area for the on-resonant Rose-Zener problem.

\emph{The case $R=1$.} At this point, starting with the initial state (\ref{IS1}), we have from Eqs.~(\ref{S2})-(\ref{S22}) and (\ref{NT2}),
\begin{eqnarray}\label{C2S2NI2}
    |C_{1}(+\infty)|^{2}&=&4A^{2},\\
    |C_{2}(+\infty)|^{2}&=&1-4A+4A^{2}.
\end{eqnarray}

It is obvious that if
\begin{eqnarray}\label{APT2}
 A=\frac{1}{2},
\end{eqnarray}
we get the complete
population inversion (CPI).

\emph{The case $R>1$.} In that case, from Eqs.~(\ref{S3})-(\ref{S33}) and (\ref{NT2}), the final populations in levels $|1\rangle$ and $|2\rangle$
 corresponding to the initial condition (\ref{IS1}) can be easily found to be
\begin{eqnarray}\label{C2S3NI2}
|C_{1}(+\infty)|^{2}&=\frac{\sinh^{2}(2A\sinh\phi)}{\sinh^{2}\phi}.\\
    |C_{2}(+\infty)|^{2}&=\frac{\sinh^{2}(\phi-2A\sinh\phi)}{\sinh^{2}\phi}.
\end{eqnarray}
If
\begin{eqnarray}\label{APT3}
2A\sinh\phi=\phi,
\end{eqnarray}
one finds that $|C_{1}(+\infty)|^{2}=1,|C_{2}(+\infty)|^{2}=0$, such that a complete population inversion (CPI) occurs.

Thus, three distinct modified pulse area conditions for CPI in the non-Hermitian two-level system (\ref{eq1}) have been identified, which are dependent on the choice of the proportional constant $R$ and given by Eqs.~(\ref{APT11}), (\ref{APT2}) and (\ref{APT3}) respectively.

\begin{figure}[htbp]
\centering
\includegraphics[width=8cm]{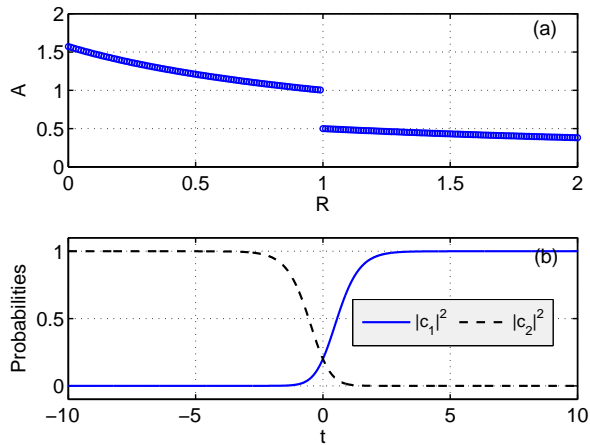}
\caption{(color online) (a) Parametric values of $(A, R)$ allowed for CPI in the system (\ref{eq1}) subjected to a synchronous modulation of the $\textrm{sech}^2$ form given by Eq.~(\ref{M2}). The allowed parameter values of $(A, R)$ are
obtained from Eqs.~(\ref{APT11})($n=0$), (\ref{APT2}) and (\ref{APT3}).
(b) Time evolution of the population distributions for the
initial state $ C_{1}(-\infty)=0,~~C_{2}(-\infty)=1$. The system parameters are set as $R=1.5$ and $2A=\textrm{arccosh} R/\sqrt{R^2-1}$.}\label{fig4}
\end{figure}

The above-mentioned CPI conditions in the parametric plane of $(A, R)$ are visualized in Fig.~\ref{fig4}(a), which are directly
obtained from Eqs.~(\ref{APT11})(the $n=0$ case), (\ref{APT2}) and (\ref{APT3}). As an example, in Fig.~\ref{fig4}(b), we select $R=1.5$ and the pulse area $2A=\textrm{arccosh} R/\sqrt{R^2-1}$ given by Eq.~(\ref{APT3}) to show the numerical results for time evolutions of the
population distributions with the initial state (\ref{IS1}). Clearly, a complete population inversion (CPI) has been fully confirmed by our numerical simulations.

\section{Conclusion}

In summary, we have presented a class of exact analytical solutions for a $\mathcal{PT}$-symmetric two-level system under synchronous combined modulations.
Such exact solutions are expressible
in terms of simple elementary functions and are applicable in the coherent control of quantum states. With these  analytical solutions,
we have demonstrated that for a cosine synchronous driving with a nonzero static component, the system behaves likes the undriven one with a phase transition from full to broken $\mathcal{PT}$ symmetry, while for a purely cosine synchronous driving, the system, although being
non-Hermitian, feature all-real quasienergy spectra for the whole range of tuning parameters. Surprisingly, we have found that coherent destruction of tunneling (CDT) and  complete
population inversion (CPI) still exist in the non-Hermitian situation. Furthermore, we have given analytically the modified pulse area conditions for such non-Hermitian analog of CPI, whose mathematical formulations are found to be dependent on the choice of the proportional constant between the two combined synchronous modulations. These simple exact analytical solutions are numerically confirmed and may find applications in any and all physical contexts
where the $\mathcal{PT}$-symmetric two-state problems arise.

The work was supported by the NSF of China under Grants 11465009,
11165009, the Doctoral Scientific Research Foundation of JingGangShan University (JZB15002), and the Program for New Century Excellent Talents in
University of Ministry of Education of China (NCET-13-0836).


\end{document}